\documentclass{article}

\usepackage{arxiv}

\usepackage[utf8]{inputenc} % allow utf-8 input
\usepackage[T1]{fontenc}    % use 8-bit T1 fonts
\usepackage{hyperref}       % hyperlinks
\usepackage{url}            % simple URL typesetting
\usepackage{booktabs}       % professional-quality tables
\usepackage{amsfonts}       % blackboard math symbols
\usepackage{nicefrac}       % compact symbols for 1/2, etc.
\usepackage{microtype}      % microtypography
\usepackage{graphicx}
\usepackage{amssymb}
\usepackage{amsmath}
\usepackage{lineno}
\usepackage{hyperref}
\usepackage{listings}
\usepackage{subfigure}
\usepackage{comment}
\usepackage{enumitem}
\usepackage[dvipsnames]{xcolor}

\title{Why POCS works, and how to make it better}

\author{
  Matteo Ravasi \\
  KAUST\\
  Thuwal, Kingdom of Saudi Arabia \\
  \texttt{matteo.ravasi@kaust.edu.sa}\\
  \And
  Nick Luiken \\
  KAUST\\
  Thuwal, Kingdom of Saudi Arabia \\
  \texttt{nicolaas.luiken@kaust.edu.sa}
  }
  
\begin{document}

\chead{Why POCS works, and how to make it better}

\maketitle

\begin{abstract}
   Projection Over Convex Sets (POCS) is one of the most widely used algorithms in geophysical data processing to interpolate seismic data. Whilst usually described as a modification of the Gerchberg-Saxton algorithm, a formal understanding of the underlying objective function and its implication for the associated optimization process is lacking to date in the literature. We show that the POCS algorithm can be interpreted as the application of the Half-Quadratic Splitting (HQS) method to the $L_0$ norm of the projected sought after data (where the projection can be any orthonormal transformation), constrained on the available traces. Similarly, the popular, apparently heuristic strategy of using a decaying threshold in POCS is revealed to be the result of the continuation strategy that HQS must employ to converge to a solution of the minimizer. Finally, we prove that for off-the-grid receivers, the POCS algorithms must operate in an inner-outer fashion (i.e., an inverse problem is solved at each outer iteration). In light of our new theoretical understanding, we suggest the use of a modern solver such as the Chambolle-Pock Primal-Dual algorithm and show that this can lead to a new POCS-like method with superior interpolation capabilities at nearly the same computational cost of the industry-standard POCS method.
\end{abstract}

\section{Introduction}
Motivated by economical and practical constraints, geophysical data acquisition campaigns typically consist of a limited portion of the sought after regular and finely sampled grids of sources and receivers. As such, reconstructing missing traces represents one of the key early steps in the geophysical processing pipeline.

In 2006, \cite{Abma20063D} proposed a method for reconstructing irregularly sampled data under the name of Projection Over Convex Sets (POCS). Owing to its intuitive nature and simplicity of implementation, POCS has become the de-facto first choice in the seismic industry. Since its introduction to the geophysical community, several works have focused on improving its effectiveness by considering different transformations that render seismic data to be sparse \cite{Yang2012Curvelet, Wang2014Dreamlet, Jicheng2018Interpolating} and relaxation strategies for the thresholding step \cite{Gao2010Irregular, Gao2012Convergence}. Recent interest has also emerged in combining the POCS iterations with neural network based transformations \cite{Park2020Seismic}. Nevertheless, a formal interpretation of POCS in the context of convex optimization is currently lacking. Whilst several authors suggest that POCS is a modified version of the Gerchberg--Saxton iterative algorithm \cite{Gerchberg1972APA}, they fail to clearly identify the underlying objective function that this method is trying to optimize. 

By leveraging recent advances in the field of convex optimization and with the help of proximal operators, we prove that the POCS iterations can be formally derived by applying the Half-Quadratic Splitting (HQS) algorithm of \cite{Geman1995Nonlinear} to the $L_0$ norm of the projected sought after data, constrained on the available traces. We also theoretically justify why the POCS algorithm benefits from applying a relaxation strategy on the thresholding parameter. After laying the mathematical foundations, we introduce two alternative solvers that can be used to solve the objective functions that underlie the POCS method. Such solvers are shown to be more flexible than HQS in that they can more naturally handle the inclusion of frequency-wavenumber masking operators and off-the-grid receivers. Numerical examples on a complex 3D synthetic dataset highlight how such solvers can lead to improved and sometimes faster convergence than the currently used POCS algorithm.

\section*{Theory}

\subsection*{Preliminaries}

Let $\mathbf{X} \in \mathbb{R}^{N_{r,y} \times N_{r,x} \times N_t}$ be a tensor containing seismic data regularly sampled along a two-dimensional spatial grid of size $N_{r,y} \times N_{r,x}$, and $\mathbf{Y} \in \mathbb{R}^{N \times N_t}$ be the physically recorded seismic data at $N$ arbitrarily selected receiver locations within the grid. Their vectorized versions, $\mathbf{x}=vec(\mathbf{X})$ and $\mathbf{y}=vec(\mathbf{Y})$, are linked via
\begin{equation}
\label{VSr3Pjurri}
\textbf{y} = \textbf{R} \textbf{x},
\end{equation}
where $\mathbf{R}\in \mathbb{R}^{N N_t \times N_{r,y}N_{r,x}N_t}$ is a \textit{restriction} operator that selects the $N$ available traces from the regularly sampled data. An alternative definition of the modelling operator is also provided, where $\tilde{\mathbf{R}} = \mathbf{R}^H \mathbf{R} \in \mathbb{R}^{N_{r,y}N_{r,x} N_t \times N_{r,y}N_{r,x}N_t}$ is a \textit{masking} operator (i.e., diagonal matrix containing 1s at the available spatial locations and 0s at the unavailable spatial locations). Similarly, we define $\tilde{\mathbf{Y}} \in \mathbb{R}^{N_{r,y} \times N_{r,x} \times N_t}$ as the recorded dataset where zero traces are interleaved to the physically recorded traces at locations of missing receivers. Its vectorized version $\tilde{\mathbf{y}}=vec(\tilde{\mathbf{Y}})$ is therefore linked to the vector $\textbf{x}$ via $\tilde{\textbf{y}} = \tilde{\textbf{R}} \textbf{x}$. Similar to the modelling operator, the following relation holds $\tilde{\textbf{y}}=\textbf{R}^H\textbf{y}$.

Finally, the \textit{proximal operator} of a possibly non-smooth function $\tau f(\textbf{x})$ is defined as:
\begin{equation}
\label{c8xnupZACY}
prox_{\tau f}(\mathbf{u}) = \underset{\mathbf{x}} {\mathrm{min}}  \; f(\mathbf{x}) + \frac{1}{2 \tau} ||\mathbf{x} - \mathbf{u}||_2^2,
\end{equation}
where $\tau > 0$. From equation \ref{c8xnupZACY} we can observe that evaluating the proximal operator involves solving an optimization problem, which compromises between minimizing the function itself and finding a solution close to $\mathbf{u}$. The introduction of the $L_2$ norm of the difference between the vector $\mathbf{u}$ and the solution $\mathbf{x}$ renders the functional strongly convex, so that is has a unique minimizer for every vector $\mathbf{u}$. As we will see in the following, proximal operators are a key component in convex optimization and closed-form solutions exist for many commonly used functions $f$.

\subsection*{The POCS algorithm}

The aim of any seismic interpolation algorithm is to reconstruct the regularly sampled data $\mathbf{X}$ from the available traces $\mathbf{Y}$. However, solving equation \ref{VSr3Pjurri} is a severely ill-posed inverse problem and requires prior knowledge about the expected solution. One way to introduce such prior knowledge is via linear sparsifying transforms, $\mathbf{S}\in \mathbb{R}^{N_{r,y}N_{r,x}N_t \times N_s}$, that provide a compact representation (i.e., few non-zero coefficients) of the regularly sampled seismic data in the transformed domain. Here $N_s$ is used to indicate the number of coefficients in the transformed domain.

The original POCS algorithm of \cite{Abma20063D} leverages a specific type of sparsifying transform, namely the multi-dimensional Fourier transform, and can be compactly expressed by the following iterations:
\begin{equation}
\label{nM9WfP5L68}
\textbf{x}^k= \tilde{\textbf{y}} + (\textbf{I}-\tilde{\textbf{R}}) \textbf{S} \mathbb{T}_\alpha( \textbf{S}^H \textbf{x}^{k-1}),
\end{equation}
where $\mathbb{T}_\alpha(x)=xH(|x| -\alpha)$ is a hard-thresholding operation controlled via the parameter $\alpha$, and $H$ is the Heaviside step function. Given a vector $\textbf{x}$, this operation acts in an element-wise fashion such that elements with an absolute value smaller than $\alpha$ are truncated to zero whilst all other elements remain untouched. The POCS algorithm owes its popularity to its simplicity of implementation and effectiveness; in fact, the main iteration of the algorithm in (\ref{nM9WfP5L68}) can be easily interpreted as follows: transform the current solution to a sparse domain of choice, apply a threshold, transform back to the original domain, extract only traces at unavailable receivers and interleave them with the true traces at available receivers.

Note that, the POCS algorithm is agnostic to the sparsifying transform $\textbf{S}$ provided that it satisfies the following conditions: $\textbf{S}^H \textbf{S}=\textbf{I}_{N_s}$ and $\textbf{S} \textbf{S}^H=\textbf{I}_{N_{r,y}N_{r,x}N_t}$ (i.e., orthogonormality) as shown below. Moreover, whilst our theory is developed for 3D seismic data, its extension to 5D data is trivial as explained in \cite{Abma2009Issues}. Finally, in practical scenarios physical receivers may be placed at locations that do not lie exactly on top of the nominal grid: the restriction operator must therefore be substituted by a spatial interpolator. The implications of such a change for the POCS algorithm will be later discussed.

\section*{Deriving POCS from HQS}

In this section, a mathematically sound derivation of the POCS iterations is provided drawing upon the theory of convex optimization. To begin with, let's define the following objective function:
\begin{equation}
\label{CjxpuQHw2w}
J=f(\textbf{x}) + g(\textbf{x}),
\end{equation}
where $f$ and $g$ are any convex functions with a known proximal operator. The HQS algorithm minimizes the functional in equatio  \ref{CjxpuQHw2w} by introducing a splitting of the form:
\begin{equation}
\label{jxDT91EJtl}
J=f(\textbf{x}) + g(\textbf{z} ) \; \text{s.t.} \; \textbf{z}=\textbf{x},
\end{equation}
and relaxing the constraint using the penalty method:
\begin{equation}
J_{HQS}=f(\textbf{x}) + g(\textbf{z} ) + \frac{\rho}{2} ||\textbf{z}-\textbf{x}||_2^2.
\end{equation}
The associated unconstrained minimization is solved in an alternating fashion:
\begin{equation}
\label{eq:hqsiters}
\begin{split}
\textbf{x}^k &= \underset{\textbf{x}}{\mathrm{argmin}}\: J_{HQS}(\textbf{x}, \textbf{z}^{k-1} ) = \underset{\textbf{x}}{\mathrm{argmin}} \; f(\textbf{x}) + \frac{\rho}{2} ||\textbf{z}^{k-1}-\textbf{x}||_2^2=prox_{\frac{1}{\rho} f}(\textbf{z}^{k-1}) \\
\textbf{z}^k &= \underset{\textbf{z}}{\mathrm{argmin}}\: J_{HQS}(\textbf{x}^k, \textbf{z})= \underset{\textbf{z}}{\mathrm{argmin}} \; g(\textbf{z})  + \frac{\rho}{2}  ||\textbf{z}-\textbf{x}^{k}||_2^2=prox_{\frac{1}{\rho} g}(\textbf{x}^{k}).
\end{split}
\end{equation}
It becomes obvious from equation \ref{eq:hqsiters} that HQS can operate with any combination of $f$ and $g$ functions provided their proximal operator can be easily computed.

Let's now define the following constrained objective function and prove that its minimization via HQS is equivalent to the POCS algorithm:
\begin{equation}
\label{clZST4gMME}
J_{POCS}= ||\textbf{S}^H\textbf{x}||_0 \; \text{s.t.} \; \textbf{y}=\textbf{R} \textbf{x},
\end{equation}
where $||.||_0$ identifies the $L_0$ norm. Note that this is equivalent to writing an unconstrained objective function of the form:
\begin{equation}
\label{RSXwiPWnum}
J_{POCS}= ||\textbf{S}^H\textbf{x}||_0 + i_{\textbf{y}=\textbf{R}\textbf{x}},
\end{equation}
where $i_{\textbf{y}=\textbf{R}\textbf{x}}$, is the indicator function of the Affine set $\{\textbf{x}\in \mathbb{R}^{N_{r,y}N_{r,x} N_t} | \textbf{y}=\textbf{R} \textbf{x}\}$. The optimization problem associated with the functional in equation \ref{clZST4gMME} can be interpreted as finding a sparse representation of the regularly sampled seismic data $\textbf{x}$ (by minimizing the $L_0$ norm in the transformed domain) whilst exactly matching the recorded data. Let $f(\textbf{x})=||\textbf{S}^H\textbf{x}||_0$ and $g(\textbf{x})=i_{\textbf{y}=\textbf{R}\textbf{x}}$, the POCS functional in equation \ref{clZST4gMME}) can be minimized via the HQS algorithm by simply repeatedly applying their proximal operators. Starting from $f$ and by leveraging the fact that $\textbf{S}$ is an orthonormal operator, the proximal operator has a closed-form solution \cite{O2014Primal},
\begin{equation}
\label{rqVF4ruiIV}
prox_{\frac{1}{\rho} f}(\textbf{x})  = \textbf{S} \mathbb{T}_{\frac{1}{\rho}}(\textbf{S}^H\textbf{x}).
\end{equation}
Similarly, since $g$ is an indicator function over an affine set, its proximal operator can be written as \cite{Parikh2014Proximal}
\begin{equation}
\label{eq:affine}
\begin{split}
prox_{\frac{1}{\rho} g}(\textbf{x}) &= \textbf{x} -
       \textbf{R}^H(\textbf{R}\textbf{R}^H)^{-1}(\textbf{R}\textbf{x}-\textbf{y}) \\
&= \textbf{x} - \textbf{R}^H(\textbf{R}\textbf{x}-\textbf{y}) \\
&= \tilde{\textbf{y}} + (\textbf{I}-\tilde{\textbf{R}})\textbf{x},
\end{split}
\end{equation}
where we have used the following property of a restriction operator, $\textbf{R}\textbf{R}^H=\textbf{I}_{N N_t}$. This proximal operator performs exactly the operation referred in the POCS literature to as the available trace re-inserting strategy. By combining equations \ref{rqVF4ruiIV} and \ref{eq:affine} together in the iterations of HQS, and taking $\alpha=1/\rho$, we obtain the iterations of the original POCS algorithm (equation \ref{nM9WfP5L68}). At this point, three key observations can be made:

\begin{itemize}
\item following the definition in \cite{Menke1991Applications}, POCS is an algorithm which seeks a solution that jointly satisfies a number of properties expressed in terms of sets (and it can be only applied to convex sets). On the other hand, the $L_0$ norm used in equation \ref{clZST4gMME} to obtain the iterations of Abma's POCS is not a convex set. We have therefore discovered that, despite its name, this algorithm is strictly-speaking not a POCS-type algorithm;
\item in order to ensure convergence to a minimizer of the original objective function, HQS must use a continuation strategy, i.e., $\rho^k \rightarrow \infty$. Looking at equation \ref{rqVF4ruiIV}, this leads to decreasing the threshold of the hard-thresholding operation during iterations. This finding explains why an effective strategy to improve the convergence of the POCS algorithm is to exponentially decrease the thresholding applied during the x-update in the transformed domain \cite{Ge2015fast};
\item several transforms commonly used in geophysics are not orthonormal. A prominent example of non-orthonormal transform is the Radon transform. In this case, equation \ref{rqVF4ruiIV} does not hold and a closed-form solution for the proximal operator does not exist. Whilst \cite{Kabir1995Restoration} use POCS-like iterations in conjunction with a parabolic Radon transform to fill missing near offsets, their algorithm does not formally solve the objective function in equation \ref{clZST4gMME}. We will see in the next section how by simply swapping HQS with another solver, the objective function in equation \ref{clZST4gMME} can still be minimized also when a non-orthonormal transform is chosen.
\end{itemize}

\section*{Modifications to the POCS algorithm}

Equipped with a solid mathematical understanding of the POCS algorithm, our next step is to define some alternative ways to minimize the functional in equation \ref{clZST4gMME}.

Two popular methods in the field of convex optimization for the minimization of such composite objective functions are the Alternating Direction Method of Multipliers (ADMM - \cite{Boyd2010Distributed}) and the Chambolle-Pock Primal-Dual algorithm (PD - \cite{Chambolle2010First}). ADMM takes a similar route to that of HQS in that it performs a splitting of the original objective function. Differently from HQS, ADMM introduces an augmented Lagrangian of the form:
\begin{equation}
\label{iUNnXkmSxS}
J_{ADMM}=f(\textbf{x}) + g(\textbf{z} ) + \textbf{y}^H(\textbf{z}-\textbf{x})+ \frac{\rho}{2} ||\textbf{z}-\textbf{x}||_2^2,
\end{equation}
where $\textbf{y}$ is the Lagrange multiplier. By solving this augmented objective function in an alternating fashion for our definition of $f$ and $g$ functions, we obtain:
\begin{equation}
\begin{split}
\textbf{x}^k &= prox_{\frac{1}{\rho} f}(\textbf{z}^{k-1}-\textbf{y}^{k-1}) =  \textbf{S} \mathbb{T}_{\frac{1}{\rho}}(\textbf{S}^H(\textbf{z}^{k-1}-\textbf{y}^{k-1}))  \\
\textbf{z}^k &= prox_{\frac{1}{\rho} g}(\textbf{x}^{k}+\textbf{y}^{k-1}) = \tilde{\textbf{y}} + (\textbf{I}-\tilde{\textbf{R}})(\textbf{x}^{k}+\textbf{y}^{k-1})\\
\mathbf{y}^{k+1} &= \mathbf{y}^{k} + \mathbf{x}^{k+1} - \mathbf{z}^{k+1}.
\end{split}
\end{equation}
The presence of a Lagrangian term equation \ref{iUNnXkmSxS} ensures that at convergence the constraint $\textbf{z}=\textbf{x}$ is verified and the original problem is solved. Theoretically, the scalar $\rho$ can therefore be kept fixed throughout iterations, relieving the user from defining a cooling strategy as required by HQS.

On the other hand, the Primal-Dual algorithm is designed to solve problems in the form of $f(\textbf{Kx})+g(\textbf{x})$ where $\textbf{K}$ is any linear operator. It proceeds finding the saddle point of the functional:
\begin{equation}
\label{Ah7bGYpiYb}
\max_{\textbf{y}}\min_{\textbf{x}} J_{PD} = \max_{\textbf{y}}\min_{\textbf{x}} \textbf{y}^H(\textbf{Kx}) + g(\textbf{x}) + f^*(\textbf{y}),
\end{equation}
where $f^*$ is the convex conjugate of $f$. This problem can be solved by the following iterations:
\begin{equation}
\label{eq:pditers}
\begin{split}
\mathbf{y}^{k+1} &= prox_{\mu f^*}(\mathbf{y}^{k} + \mu \mathbf{K}\bar{\mathbf{x}}^{k})\\
\mathbf{x}^{k+1} &= prox_{\tau g}(\mathbf{x}^{k} - \tau \mathbf{K}^H \mathbf{y}^{k+1}) \\
\bar{\mathbf{x}}^{k+1} &= 2 \mathbf{x}^{k+1} - \mathbf{x}^k.
\end{split}
\end{equation}
where $\tau$ and $\mu$ represent the step-lengths of the sub-gradients of $f$ and $g$. To ensure convergence $\tau \mu L^2 < 1$, where $L^2 = ||\textbf{K}||_2^2 = \lambda_{max}(\mathbf{K}^H\mathbf{K})$ is the spectral radius of the linear operator $\textbf{K}$.

If we choose $\textbf{K}= \textbf{I}$, $f(\textbf{x})=||\textbf{S}^H\textbf{x}||_0$, and $g(\textbf{x})=i_{\textbf{y}=\textbf{R}\textbf{x}}$, the iterations in equation \ref{eq:pditers} can be employed to optimize the functional in equation \ref{clZST4gMME}. To evaluate step 1 in equation \ref{eq:pditers}, the proximal operator of $f^*$ is required: this can be simply obtained from the proximal operator of the original function $f$ via the Moreau identity \cite{Parikh2014Proximal}, $prox_{\mu f^*}(\mathbf{x})=\mathbf{x} -\frac{1}{\mu}prox_{\frac{f}{\mu}}(\frac{\mathbf{x}}{\mu})$. All other expression for the proximal operators are the same as those provided in equations \ref{rqVF4ruiIV} and \ref{eq:affine}, with the only difference that the inputs in the PD iterations vary from those of HQS and ADMM. However, a different choice that can lead to increased reconstruction capabilities is made here: more specifically, by defining $\textbf{K}= \textbf{S}^H$, $f(\textbf{x})=||\textbf{x}||_0$ and $g(\textbf{x})=i_{\textbf{y}=\textbf{R}\textbf{x}}$, the functional in equation \ref{clZST4gMME} is again obtained and solved via the Primal-Dual algorithm. Similarly, the ADMM algorithm can be adapted to such a scenario: the resulting algorithm takes the name of Linearized-ADMM (L-ADMM) as described in detail in \cite{Parikh2014Proximal}. Nevertheless, as previously reported in both the context of natural image reconstruction problems \cite{Heide2014FlexISP} and seismic inversion \cite{Ravasi2021joint}, our numerical examples will show the superior capabilities of PD against L-ADMM.

Since the transform $\textbf{S}$ is moved out of the function $f$, the Primal-Dual algorithm can be equally applied to equation \ref{clZST4gMME} with orthogonal and non-orthogonal sparsifying transforms. This represents a benefit compared to the HQS formulation. Another benefit of not requiring $\textbf{K}$ to be orthogonal is represented by the fact that a masking operator $\textbf{M}_{fk}$ operating in the frequency-wavenumber can be included to enforce the spectrum of the reconstructed data to lie inside the expected signal cone (i.e., erase energy from the aliasing regions). We can define $\textbf{K}=\textbf{M}_{fk} \textbf{S}^H$ such that the problem to solve becomes:
\begin{equation}
\label{VP3yg6GPD9}
J_{Masked-POCS}= ||\textbf{M}_{fk}\textbf{S}^H\textbf{x}||_0 \; \text{s.t.} \; \textbf{y}=\textbf{R} \textbf{x}.
\end{equation}
We note that the idea of including a masking operation within the original POCS iterations in order to handle regularly sampled data was originally proposed by \cite{Zhang2020anti}. However, since the resulting function $f$ does not have a closed-form proximal operator, their iterative scheme is not guaranteed to solve the objective function in equation \ref{VP3yg6GPD9}.

%Finally, various publications have suggested that the POCS algorithm can also be used to jointly reconstruct and denoise seismic data (e.g., \cite{Gao2010Irregular} \cite{Wang2016Improved}). This is simply accomplished by weighting the contribution of the true and reconstructed traces at available locations. Whilst we struggle to find an objective function that could formally explain this modified algorithm, a more rigorous approach in the presence of unwanted noise in the data is to substitute the Affine set indicator in (\ref{clZST4gMME}) with a Euclidean ball, i.e., $||\textbf{y} -\textbf{R} \textbf{x}||_2<\sigma$ where $\sigma$ represents the standard deviation of the noise in the data. Unfortunately, the proximal operator of this function does not have a closed-form solution and, therefore, the modified objective function cannot be easily solved via HQS. On the other hand, by defining $\textbf{K}=\textbf{R}$, $f(\textbf{x})=||\textbf{S}^H\textbf{x}||_0$ and $g(\textbf{x})=||\textbf{y} -\textbf{x}||_2<\sigma$, a solution to this problem can be easily found via PD (or L-ADMM) using the following expression for the proximal operator of the Euclidean ball:

%\begin{equation}
%\label{e6KZ1h3sQi}
%prox_{\frac{1}{\rho} g}(\textbf{x}) = \mathbf{y} + \frac{\sigma}
%        {max\{ ||\mathbf{x} - \mathbf{y}||_2^2, \sigma \}}(\mathbf{x} - \mathbf{y}).
%\end{equation}

\section*{POCS with off-the-grid receivers}

In this final section we discuss the implications of off-the-grid receivers for the POCS algorithm. As briefly mentioned in the Introduction section, the restriction operator $\textbf{R}$ in equation \ref{VSr3Pjurri} is not suitable for scenarios where physical receivers do not lie exactly on on the nominal recording grid. A bilinear (or sinc) interpolation operator $\textbf{B}\in R^{N N_t \times N_{r,y}N_{r,x}N_t}$ is therefore required to handle such a scenario. Contrary to the restriction operator, this operator does not satisfy the condition $\textbf{B} \textbf{B}^H=\textbf{I}_{N N_t}$ required by an Affine set to have a closed-form solution for its proximal operator (equation \ref{eq:affine}). As a consequence, for each proximal evaluation, the problem $(\textbf{B} \textbf{B}^H)\textbf{z}=\textbf{B}\textbf{x} -\textbf{y}$ must be solved. This can be accomplished using a gradient-based iterative solver (e.g., CGLS, LSQR). Finally, the evaluation of the proximal operator reduces to $prox_{\frac{1}{\rho} g}(\textbf{x}) = \textbf{x} - \textbf{B}^H\textbf{z}$. Although an approximate solution is usually enough and therefore a few iterations suffice, this slightly increases the cost of the overall reconstruction process. We note that this differs from what has been previously suggested in the literature by \cite{Jiang2017Compressive}: their extended POCS (EPOCS) method ignores the $\textbf{B} \textbf{B}^H$ term, which can however deviate from an identity matrix as later shown in the Numerical Results section.

\section*{Numerical Results}

\begin{figure}[!htbp]
\centering
\includegraphics[width=1\linewidth]{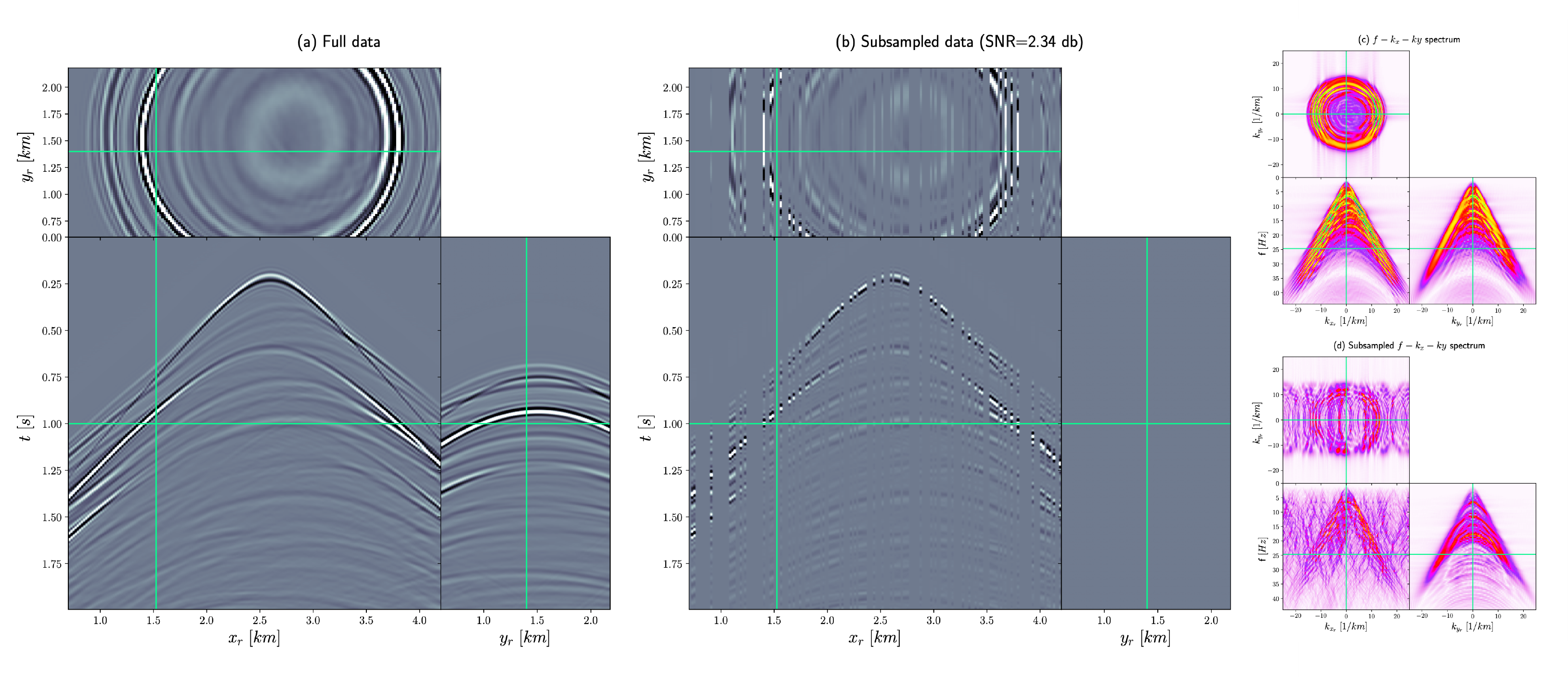}
\caption{a) Full data, b) Decimated data obtained by selecting 40\% of receiver lines along the X-axis. c-d) Frequency-wavenumber spectra of the full and decimated data, respectively.}
\label{fig1}
\end{figure}

\begin{figure}[!htbp]
\centering
\includegraphics[width=1\linewidth]{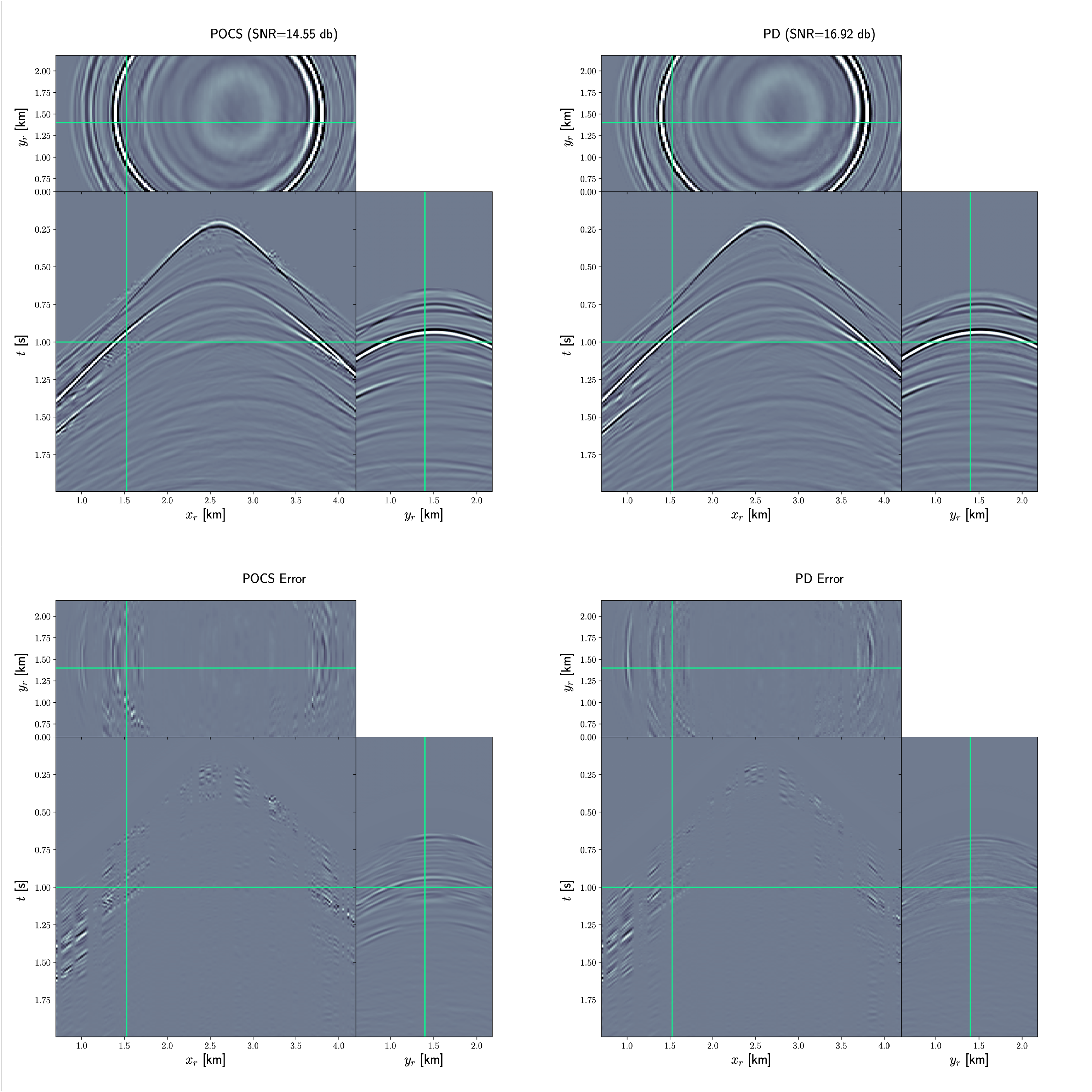}
\caption{Reconstruction with a) POCS algorithm with exponential decaying threshold, b) PD algorithm with constant threshold. c-d) Corresponding reconstruction errors.}
\label{fig2}
\end{figure}

\begin{figure}[!htbp]
\centering
\includegraphics[width=1\linewidth]{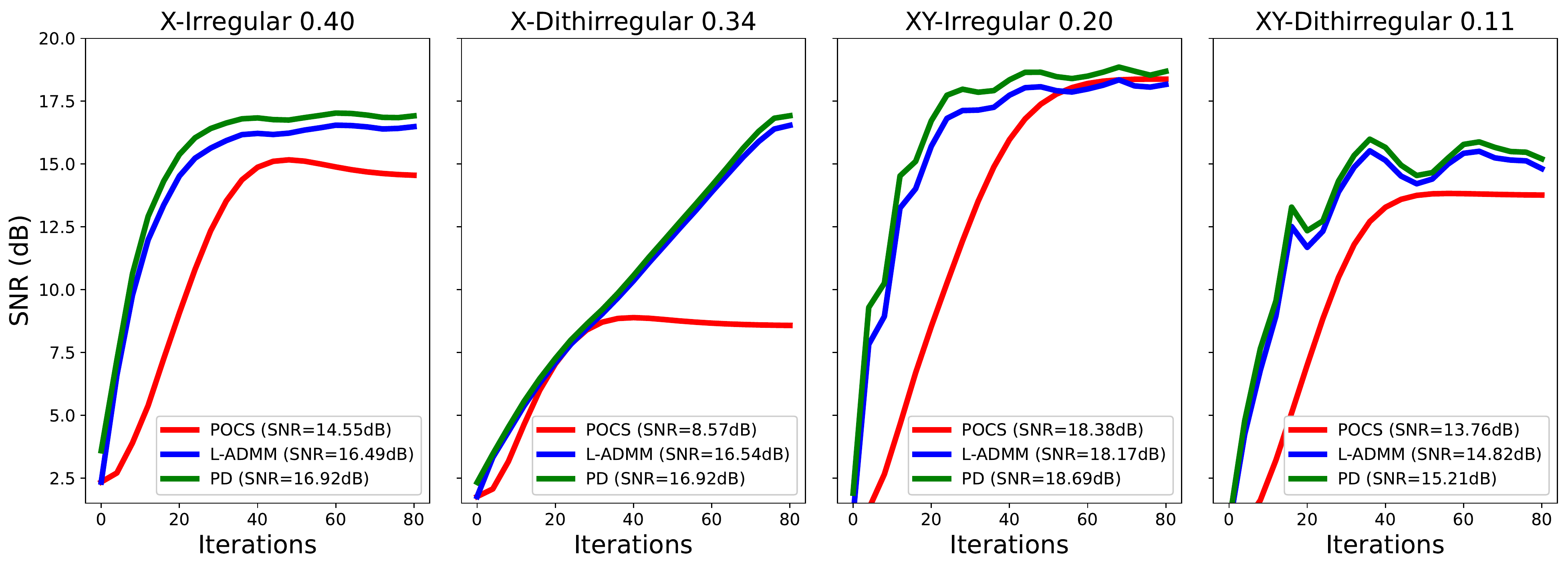}
\caption{SNRs as function of iterations for four different subsampling schemes.}
\label{fig3}
\end{figure}

The original POCS algorithm and the proposed modifications are now applied to a synthetic 3D seismic dataset modeled using a modified version of the SEG/EAGE Salt and Overthrust Model to which a water column of 300m has been added on top. The dataset is created using an acoustic finite-difference modelling code with free-surface activated and a source signature with flat spectrum between 3 and 45 Hz. The source is placed at a depth of 10m in the middle of the model and receivers are placed along the seafloor in a regular grid of size $176 \times 80$ with spacing $dx_R=20m$ and $dy_R=20m$. The total recording time is 2sec. Figure~\ref{fig1} displays the regularly sampled shot gather and a decimated version obtained by randomly selecting 40\% of the receiver lines along the X-axis. The reconstruction produced after 80 iterations of the original POCS algorithm with the exponential thresholding decay proposed in \cite{Gao2010Irregular} is shown in Figure~\ref{fig2}a. The reconstruction obtained from the Primal-Dual algorithm with constant thresholding is displayed in Figure~\ref{fig2}b. In both cases, interpolation is performed in small spatio-temporal overlapping patches of size $32 \times 32 \times 32$ with an overlap of $8 \times 8 \times 6$. Moreover, the seismic shot gather is time-shifted by $t(y_R, x_R)=[(y_S -y_R)^2 + (x_S -x_R)^2 + (z_S -z_R)]^p/v_{sea} + t_{ZO, off}$, with $p=0.43$, $v_{sea}=1500m/s$ and $t_{ZO, off}=50ms$, in order to reduce aliasing prior to reconstruction and shifted back afterwards. Figure~\ref{fig3}a displays the signal-to-noise (SNR) over iterations for the two methods displayed in Figure~\ref{fig2} as well as for the L-ADMM algorithm (blue line). Moreover, the final SNRs are reported in the legends of Figure~\ref{fig3}. Similarly, SNR curves are displayed for three other subsampling strategies, namely dithered subsampling \cite{Hennenfent2008Simply} along the X-axis, irregular sampling along both the X- and Y-axis, and dithered subsampling along both axes. In this figure, we can observe two different behaviors: when subsampling is performed along the X-axis, PD both converges faster than HQS and the overall solution is also improved. On the other hand, when subsampling is performed in a fully random fashion along the entire grid of receivers, the quality of the seismic data reconstructed using the new PD algorithm is more in line with that of the original POCS algorithm, however faster convergence is retained. Note that whilst creating fully random subsampled data is known to be obtained from a compressive sensing point of view \cite{Candes2008Introduction}, applying subsampling along a single spatial direction is much easier to achieve in practical applications: the clear improvement of our PD-POCS algorithm over the original POCS algorithm is therefore of significant relevance for the seismic and geophysical industry.

\begin{figure}[!htbp]
\centering
\includegraphics[width=1\linewidth]{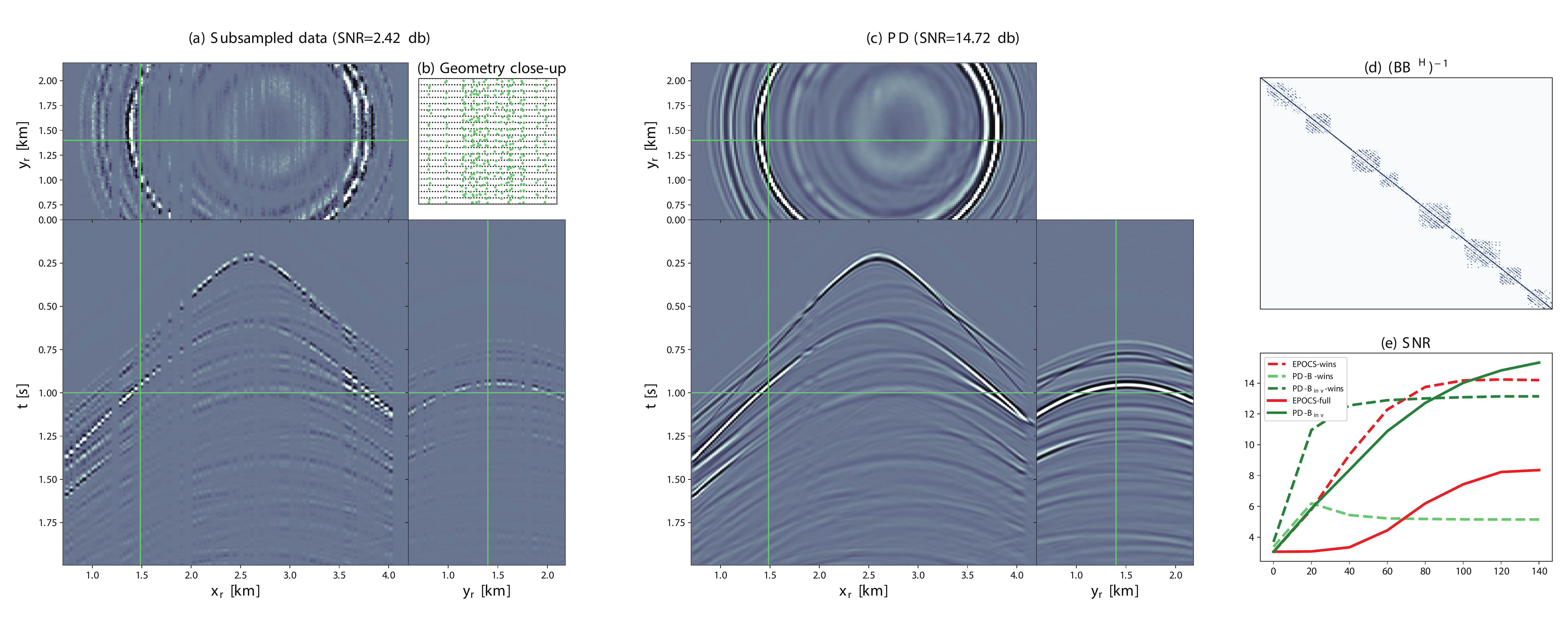}
\caption{Off-the-grid interpolation example. a) Subsampled data, b) Geometry (black dots: regular grid, green dots: available receivers), c) PD reconstruction, d) $(\mathbf{BB}^H)^{ -1}$ matrix, and e) SNRs a function of iterations.}
\label{fig4}
\end{figure}

As a second example, we consider the problem of off-the-grid receivers. We use the same synthetic dataset and create a geometry that contains 60\% of the original receivers (Figure~\ref{fig4}a); however, in this case, their position does not fall exactly on the nominal grid (see Figure~\ref{fig4}b). The reconstructed wavefield using the PD method and 2 iterations of LSQR for the evaluation of the proximal operator of $g$ is displayed in Figure~\ref{fig4}c. The corresponding SNR as a function of iterations is compared in Figure~\ref{fig4}e with that of the EPOCS algorithm and the PD method with inaccurate evaluation of the proximal of $g$, i.e. $(\textbf{B} \textbf{B}^H)^{ -1}\approx\mathbf{I}$. We observe that whilst EPOCS seems to be less affected than PD by such an approximation, the proposed method outperforms it both in terms of convergence speed and overall quality of reconstruction. This result highlights the importance of correctly handling the off-diagonal elements in $(\textbf{B} \textbf{B}^H)^{ -1}$(Figure~\ref{fig4}d).

\section*{Conclusion}
We have presented new insights into the mathematical foundations of the POCS algorithm based on the theory of proximal operators. The aim of this work is two-fold: for the math savvy readers, our derivation provides a theoretically sound explanation for why POCS works in the context on seismic interpolation. The POCS iterations come essentially from a HQS-based minimization of the $L_0$ norm of the reconstructed seismic data in a transformed domain of choice, constrained on the actual spatial-temporal measurements. The connection with the HQS method also explains why better convergence is observed when employing a varying thresholding as a function of iterations: this is nothing more than employing the continuation strategy, which is strictly required by HQS to successfully converge. For the practitioners, we have suggested a number of changes to the underlying objective function and optimizer that can be easily implemented in existing codebases; our numerical example shows how such changes can increase the speed of convergence as well as the overall quality of the reconstructed wavefield and better handle the case of off-the-grid receivers.

\section*{Acknowledgements}
The authors thank KAUST for supporting this research. All numerical examples have been produced using the PyLops library \cite{Ravasi2020PyLopsA} and are available at \url{https://github.com/DIG-Kaust/PyPOCS}.

\bibliographystyle{unsrt}  
\bibliography{references}

\end{document}